\documentclass{iau} 
\usepackage{graphicx}

\title[Origin of radio pulsar polarisation] 
{Coherent origin of peculiar polarization \\ in radio pulsars}

\author[Jaros{\l}aw Dyks]   
{Jaros{\l}aw Dyks}

\affiliation{Nicolaus Copernicus Astronomical Center, Bartycka 18, 00-716 Warsaw, Poland
\\ email: {\tt jinx@ncac.torun.pl}} 

\pubyear{2017}
\volume{337}  
\jname{Pulsar Astrophysics -­ The Next 50 Years}
\editors{P. Weltevrede, B.B.P. Perera, L. Levin Preston \& S. Sanidas, eds.}

\begin{document}

\maketitle

\begin{abstract}
The observed polarization of radio pulsars involves 
several peculiar effects, such as comparable amount of orthogonal 
polarization modes (OPMs) which often 
bear the same handedness of circular polarisation $V$. 
In the average profiles of B1913$+$16 and B1933$+$16,  
orthogonal jumps of polarization angle (PA) are observed to occur 
at the maximum $V$, instead of $V=0$. 
High levels of $V$ are also observed in core components (eg.~in B1237$+$25),
where they are accompanied by strong distortions of PA from 
the rotating vector model.  In weakly polarized emission, 
PA jumps by $45^\circ$ are observed in B1919$+$21 and B0823$+$26.
It is shown that all these peculiarities can be interpreted in a model which 
assumes coherent addition of waves in natural propagation modes. 
\end{abstract}

\firstsection 
\section{The nature of orthogonal polarization modes}

The model assumes that pulsars emit a linearly polarized signal 
$\vec E$ that enters some intervening birefringent region, characterised 
by the polarization basis $(\vec x_1, \vec x_2)$ (Fig.~1a). The basis 
is misaligned with $\vec E$ by the incident angle $\psi_{\rm in}$. 
The incident signal is split into two natural mode waves $\vec E_1$ (dotted) 
and $\vec E_2$ (solid) that are orthogonally polarized with respect to each
other.  
While propagating through the region, the waves acquire a
relative phase lag $\Delta\phi$ (Fig.~1a). After leaving the region, the
waves combine, i.e.~are added coherently. Thus, the model is empirical 
and similar to the Faraday rotation effect.  

The two main parameters, 
$\psi_{\rm in}$ and $\Delta\phi$, are drawn from  
statistical distributions $N_{\psi, \rm in}$ and $N_{\Delta\phi}$, 
supposedly produced by the stochastic nature 
of the emission and propagation processes. The distributions represent 
the spread of values as recorded in single pulse observations
at a fixed pulse longitude.
As described in Dyks (2017, hereafter D17),
the peak of $N_{\psi, \rm in}$ is determined by the relative
orientation of the magnetic field in the emission and intervening regions. 
The peak position and width of $N_{\Delta\phi}$ are treated as free
parameters. In general, arbitrary pairs of $\psi_{\rm in}$ 
and $\Delta\phi$ produce 
elliptically polarized radiation, with the ellipse major axis, hence the PA, 
 at any orientation with respect to the main polarisation directions 
$\vec x_1$ and $\vec x_2$. However, when 
the lag distribution $N_{\Delta\phi}$ encompasses 
the value of $\Delta\phi=\pi/2$, the situation is different.
As shown in Fig.~1, for $\Delta\phi=\pi/2$, only the ellipses parallel to
either $\vec x_1$ or $\vec x_2$ are produced, regardless of the 
$\psi_{\rm in}$ value. Note that in Fig.~1 the 
same phase lag of $\pi/2$ is applied to two different incident
angles: $\psi_{\rm in}=65^\circ$ (Fig.~1a) and $25^\circ$ (1b).
In this way two observed orthogonal modes are produced, 
as represented by the ellipses $\rm M_1$ and $\rm M_2$. Importantly, 
these observed modes
should be discerned from the natural orthogonal 
propagation modes that are presented by the waves $\rm m_1$ and $\rm m_2$. 
The coherently produced orthogonal modes ($\rm M_1$ and $\rm M_2$) 
are elliptical, and can easily have the same handedness of $V$, which is 
the case for the values of $\psi_{\rm in}$ used in Fig.~1. 

For phase lags in the vicinity of $\pi/2$, the orientation of the modal
ellipses is always close to $\vec x_1$ or $\vec x_2$, so the corresponding
(modal) PAs vastly outnumber other PA values (see Fig.~4 in D17). 
Here `other values' means PAs for $\Delta\phi\ne\pi/2$ 
and arbitrary $\psi_{\rm in}$ that correspond to    
ellipses misaligned with respect to $\vec x_1$ or $\vec x_2$. These nonmodal
PA values contribute to wide PA distributions that accompany the narrow
modal PA peaks (in a PA distribution observed at a given pulse longitude).


\begin{figure}[t]
\begin{center}
 \includegraphics[width=5in]{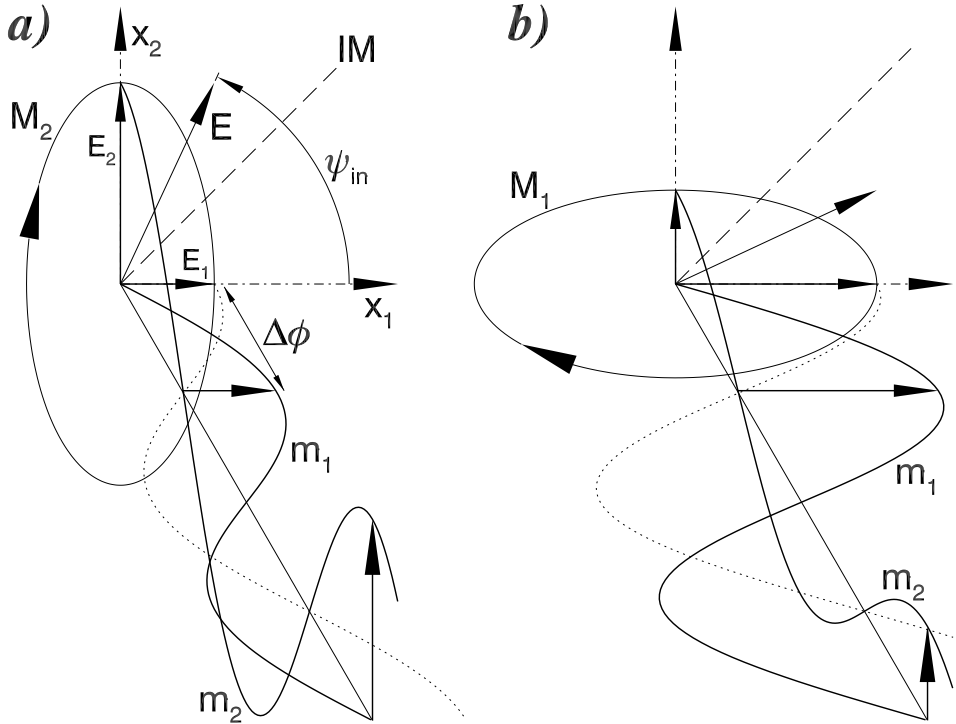} 
 \caption{The origin of observed orthogonal polarization modes, 
represented by the
 ellipses $M_1$ and $M_2$, 
as a coherent sum of the phase lagged natural mode waves $m_1$ and $m_2$. 
Note the same handedness of both modes.}
   \label{fig1}
\end{center}
\end{figure}

In the course of pulsar rotation, $\psi_{\rm in}$ is changing with 
the pulse longitude. 
When $\vec E$ coincides with the dashed intermode separatrix 
(i.e.~$\psi_{\rm in}=45^\circ$), the orthogonal mode jump at maximum $V$
occurs. The maximum degree of circular polarisation that is observed 
at the PA jump, 
 depends on the peak position and width of $N_{\Delta\phi}$. 
The step-wise PA curve of B1913$+$16 (Weisberg \& Taylor 2002), 
with the OPM transitions
at peaks of $V$, can be interpreted in this way (cf.~figs 1 and 7 in D17). 
The application of the model to the loop-shaped core PA distortions 
(Mitra et al.~2016),
and to the $45^\circ$ PA jump, is also described in D17.
\nocite{wt02, mra16, d17}

The research was funded by the National Science Center grant
DEC-2011/02/A/ST9/ 00256. Participation in the conference was covered by
NCAC.



\end{document}